\documentclass{sig-alternate-05-2015}

\usepackage{graphicx}
\usepackage[usenames]{color}
\usepackage{booktabs}
\usepackage{comment}
\usepackage{ifthen}
\usepackage{booktabs}
\usepackage{array}
\usepackage{subfigure}
\usepackage{framed}
\usepackage{xspace}
\usepackage{verbatim}
\usepackage{enumitem}

\definecolor{gray}{RGB}{90,90,90}
\newcommand\sbar[2]{{\color{gray}\rule{\dimexpr 1cm * #1 / #2}{6pt}}}

\newboolean{showcomments}
\setboolean{showcomments}{true}
\ifthenelse{\boolean{showcomments}}
{ 
	\newcommand{\nb}[2]{
		\fbox{\bfseries\sffamily\scriptsize#1}
		{\sf\small$\blacktriangleright$\textit{#2}$\blacktriangleleft$}
	}
}
{ 
	\newcommand{\nb}[2]{}
}

\newcommand{\mcode}[1]{$\tt #1$} 
\newcommand{\quotes}[1]{``#1''}

\newcommand{\margin}[0]{\\[-0.3cm]}

\newcommand{\totalProjects}[0]{748\xspace}
\newcommand{\activeProjects}[0]{471\xspace}
\newcommand{\activeProjectsPercentage}[0]{63\%}
\newcommand{\refactoredProjects}[0]{285\xspace}

\newcommand{\studiedProjects}[0]{124\xspace}
\newcommand{\studiedProjectsPercentage}[0]{17\%}
\newcommand{\studiedDevelopers}[0]{222\xspace}
\newcommand{\toolName}[0]{RefactoringMiner\xspace}
\newcommand{\refactoringTypes}[0]{12\xspace}
\newcommand{\totalMotivationThemes}[0]{44\xspace}
\newcommand{\detectedRefactorings}[0]{2,241\xspace}
\newcommand{\studiedRefactorings}[0]{463\xspace}
\newcommand{\commitsWithTruePositiveRefactoring}[0]{539\xspace}
\newcommand{\commitsWithDetectedRefactoring}[0]{729\xspace}
\newcommand{\commitsWithRefactoringAndResponse}[0]{222\xspace} 

\begin{document}

\setcopyright{acmcopyright}

\toappear{}
%

\title{Why We Refactor? Confessions of GitHub Contributors}

%
%
%
%
%

\numberofauthors{3} 
%
\author{
%
%
\alignauthor
Danilo Silva\\
       \affaddr{Universidade Federal de Minas Gerais, Brazil}\\
       \email{danilofs@dcc.ufmg.br}
\alignauthor
Nikolaos Tsantalis\\
       \affaddr{Concordia University}\\
       \affaddr{Montreal, Canada}\\
       \email{tsantalis@cse.concordia.ca}
\alignauthor Marco Tulio Valente\\
       \affaddr{Universidade Federal de Minas Gerais, Brazil}\\
       \email{mtov@dcc.ufmg.br}
}
\date{5 August 2015}

\maketitle
\begin{abstract}
Refactoring is a widespread practice that helps developers to improve the maintainability and readability of their code.
However, there is a limited number of studies empirically investigating the actual motivations behind specific refactoring operations
applied by developers.
To fill this gap, we monitored Java projects hosted on GitHub to detect recently applied refactorings,
and asked the developers to explain the reasons behind their decision to refactor the code.
By applying thematic analysis on the collected responses,
we compiled a catalogue of \totalMotivationThemes distinct motivations for \refactoringTypes well-known refactoring types. 
We found that refactoring activity is mainly driven by changes in the requirements and much less by code smells.
{\sc Extract Method} is the most versatile refactoring operation serving 11 different purposes.
Finally, we found evidence that the IDE used by the developers affects the adoption of automated refactoring tools.
\end{abstract}

%
%

%
%

%
%


\section{Introduction}


Refactoring is the process of improving the design of an existing code base, 
without changing its behavior~\cite{Opdyke:1992}. Since the beginning, the adoption of refactoring
practices was fostered by the availability of refactoring catalogues, as the 
one proposed by Fowler~\cite{Fowler:1999}.
These catalogues define a name and describe the mechanics of each refactoring,
as well as demonstrate its application through some code examples.
They also provide a {\em motivation} for the refactoring,
which is usually associated to the resolution of a code smell. For example, {\sc Extract Method} is
recommended to decompose a large and complex method or to eliminate code
duplication. As a second example, {\sc Move Method} is associated to smells like Feature Envy and Shotgun Surgery~\cite{Fowler:1999}.

There is a limited number of studies investigating the real motivations driving the refactoring practice based on interviews and feedback from actual developers.
Kim et al.~\cite{kim-tse-2014} explicitly asked developers ``in which situations do you perform refactorings?'' and recorded 10 code symptoms that motivate developers to initiate refactoring.
Wang~\cite{Wang:2009} interviewed professional software developers about the
major factors that motivate their refactoring activities and recorded human and social factors affecting the refactoring practice.
However, both studies were based on general-purpose surveys or interviews that were not focusing the discussion on specific refactoring operations applied by the developers, but rather on general opinions about the practice of refactoring.

\noindent\textbf{Contribution}:
To the best of our knowledge, this is the first study investigating \textit{the motivations behind refactoring based on the actual explanations of developers on specific refactorings they have recently applied}.
To fill this gap on the empirical research in this area, 
we report a large scale study centered on \studiedRefactorings refactorings identified in \commitsWithRefactoringAndResponse commits  
from \studiedProjects popular, Java-based projects hosted on GitHub. In this study,
we asked the developers who actually performed these refactorings to explain the reasons 
behind their decision to refactor the code.
Next, by applying thematic analysis~\cite{Cruzes:2011}, we categorized their responses into different themes of motivations. 
Another contribution of this study is that we make publicly available\footnote{http://aserg-ufmg.github.io/why-we-refactor} the data collected and the tools used to enable the replication of our findings and facilitate future research on refactoring.



\noindent\textbf{Relevance to existing research}: The results of this empirical study are important for two main reasons.

First, having a list of motivations driving the application of refactorings can help researchers and practitioners
to infer rules for the automatic detection of these motivations when analyzing the commit history of a project.
Recent research has devised techniques to help in understanding better the practice of code evolution by
identifying frequent code change patterns from a fine-grained sequence of code changes~\cite{Negara:2014},
isolating non-essential changes in commits~\cite{Kawrykow:2011}, and
untangling commits with bundled changes (e.g., bug fix and refactoring)~\cite{Dias:2015}.
In addition, we have empirical evidence that developers tend to interleave refactoring with other types of programming activity~\cite{MurphyHill2012}, i.e., developers tend to \textit{floss refactor}.
Therefore, \textit{knowing the motivation behind a refactoring can help us to understand better other related changes in a commit}.
In fact, in this study we found several cases where developers extract methods in order to make easier the implementation of a feature or a bug fix.

Second, having a list of motivations driving the application of refactorings can help researchers and practitioners
to develop refactoring recommendation systems tailored to the actual needs and practices of the developers.
Refactoring serves multiple purposes~\cite{Fowler:1999}, such as improving the design, understanding the code~\cite{DuBois:2005}, finding bugs, and improving productivity.
However, research on refactoring recommendation systems~\cite{Bavota:2015} has mostly focused on the design improvement aspect of refactoring by
proposing solutions oriented to code smell resolution.
For example, most refactoring recommenders have been designed based on the concept that developers extract methods either to eliminate code duplication, or decompose long methods~\cite{Tsantalis:2011, Silva:2014, Tairas:2012, Hotta:2012, Meng:2015, Tsantalis:2015}.
In this study, we found 11 different reasons behind the application of {\sc Extract Method} refactorings.
Each motivation requires a different strategy in order to detect suitable refactoring opportunities.
Building refactoring recommendation systems tailored to the real needs of developers will help to promote more effectively the practice of refactoring to the developers,
by recommending refactorings helping to solve the actual problems they are facing in maintenance tasks.

\section{Related Work}

Refactoring is recognized as a fundamental practice to maintain a healthy code base~\cite{Fowler:1999,Beck:1999,Opdyke:1992,mens:survey:2004}.
For this reason, vast empirical research was recently conducted to extend our knowledge on this practice.

\noindent\textbf{Studies on refactoring practices}:
Murphy et al.~\cite{Murphy:2006} record the first results on refactoring usage, collected using the Mylar Monitor, a standalone framework that collects and reports trace information about a user's activity in Eclipse.
Murphy-Hill et al.~\cite{MurphyHill2012} rely on multiple data sources to reveal how developers practice refactoring activities.
They investigate nine hypotheses about refactoring usage and conclude for instance that commit messages do not reliably indicate the presence of refactoring,
that programmers usually perform several refactorings within a short time period, and that 90\% of refactorings are performed manually.
Negara et al.~\cite{negara2013} provide a detailed breakdown on the manual and automated usage of refactoring, using a large corpus of refactoring instances detected using an algorithm that infers refactorings from fine-grained code edits.
As their central findings, they report that more than half of the refactorings are performed manually and that 30\% of the applied refactorings do not reach the version control system.

\noindent\textbf{Studies based on surveys \& interviews}:
Kim et al.~\cite{Kim:2012:FSE, kim-tse-2014} present a field study of refactoring benefits and challenges in a major software organization.
They conduct a survey with developers at Microsoft regarding the cost and risks of refactoring in general, and the adequacy of refactoring tool support,
and find that the developers put less emphasis on the behavior preserving requirement of refactoring definitions.
They also interview a designated Windows refactoring team to get insights into how system-wide refactoring was carried out,
and report that the binary modules refactored by the refactoring team had a significant reduction in the number of inter-module dependencies and post-release defects.
Wang~\cite{Wang:2009} interviews 10 professional software developers and finds a list of intrinsic (i.e., self-motivated) and external (i.e., forced by peers or the management) factors motivating refactoring activity.


\noindent\textbf{Studies on refactoring tools}:
Vakilian et al.~\cite{Vakilian:2012} reveal many factors that affect the appropriate and inappropriate use of refactoring tools.
They show for example that novice developers may underuse some refactoring tools due to lack of awareness.
Murphy-Hill et al.~\cite{Murphy-Hill:2008} investigate the barriers in using the tool support provided for the {\sc Extract Method} refactoring~\cite{Fowler:1999}.
They report that users frequently made mistakes in selecting the code fragment they want to extract and that error messages from refactoring engines are hard to understand.
Murphy-Hill et al.~\cite{MurphyHill2012} show that 90\% of configuration defaults in refactoring tools are not changed by the developers.
As a practical consequence of these studies, refactoring recommendation systems~\cite{Bavota:2015} have been proposed to foster the use of refactoring tools and leverage the benefits of refactoring,
by alerting developers about potential refactoring opportunities~\cite{Tsantalis:2009, Tsantalis:2011, Bavota:2011, Sales:2013, Bavota:2014, Silva:2014}. 

\noindent\textbf{Studies on refactoring risks}:
Kim et al.~\cite{Kim:2011} show that there is an increase in the number of bug fixes after API-level refactorings.
Rachatasumrit and Kim~\cite{Kim:2012:testing} show that refactorings are involved in almost half of the failed test cases.
Wei{\ss}gerber and Diehl show that refactorings are sometimes followed by an increasing ratio of bug reports~\cite{Weissgerber:2006}.

However, existing studies on refactoring practices do not investigate in-depth the {\em motivation} behind specific refactoring types, i.e., why developers decide to perform a certain refactoring operation.
For instance, Kim et al.~\cite{kim-tse-2014} do not differentiate the motivations between different refactoring types,
and Wang~\cite{Wang:2009} does not focus on the technical motivations, but rather on the human and social factors affecting the refactoring practice in general.
The only exception is a study conducted by Tsantalis et al.~\cite{tsantalis_empiricalstudy},
in which the authors themselves manually inspected the relevant source code before and after the application of a refactoring with a text diff tool, to reveal possible motivations for the applied refactorings.
Because they conducted this study without asking the opinion of the developers who actually performed the refactorings,
the interpretation of the motivation can be considered subjective and biased by the opinions and perspectives of the authors.
In addition, the manual inspection of source code changes is a rather tedious and error-prone task that could affect the correctness of their findings.
Finally, the examined refactorings were collected from the history of only three open source projects, which were libraries or frameworks.
This is a threat to the external validity of the study limiting the ability to generalize its findings beyond the characteristics of the selected projects.
In this study, we collected refactorings from \studiedProjects different projects, and asked the developers who actually performed these refactorings to explain the reasons behind their decision to refactor the code.

\section{Research Methodology}
\label{SecMethodology}




\subsection{Selection of GitHub Repositories}

First, we selected the top 1,000 Java repositories ordered by popularity in GitHub (stargazers count) that are not forks. From this initial list, we discarded the lower quartile ordered by number of commits, to focus the study on repositories with more maintenance activity. The final selection consists of \totalProjects repositories, including well-known projects, like {\sc JetBrains/intellij-community}, {\sc apache/cassandra}, {\sc e\-lastic/elasticsearch}, {\sc gwtproject/gwt}, and {\sc spring-projects/spring-framework}.

\begin{figure*}[!t]
\subfigure[ref1][Age]{\includegraphics[width=0.232\linewidth]{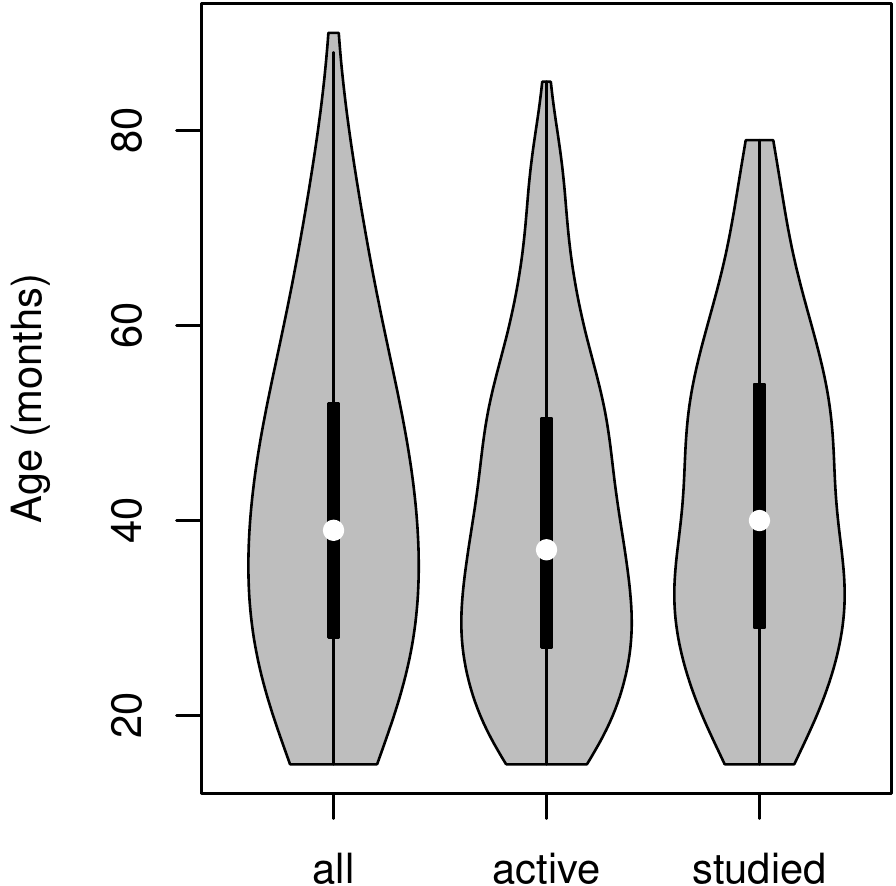}}
\quad
\subfigure[ref2][Commits]{\includegraphics[width=0.232\linewidth]{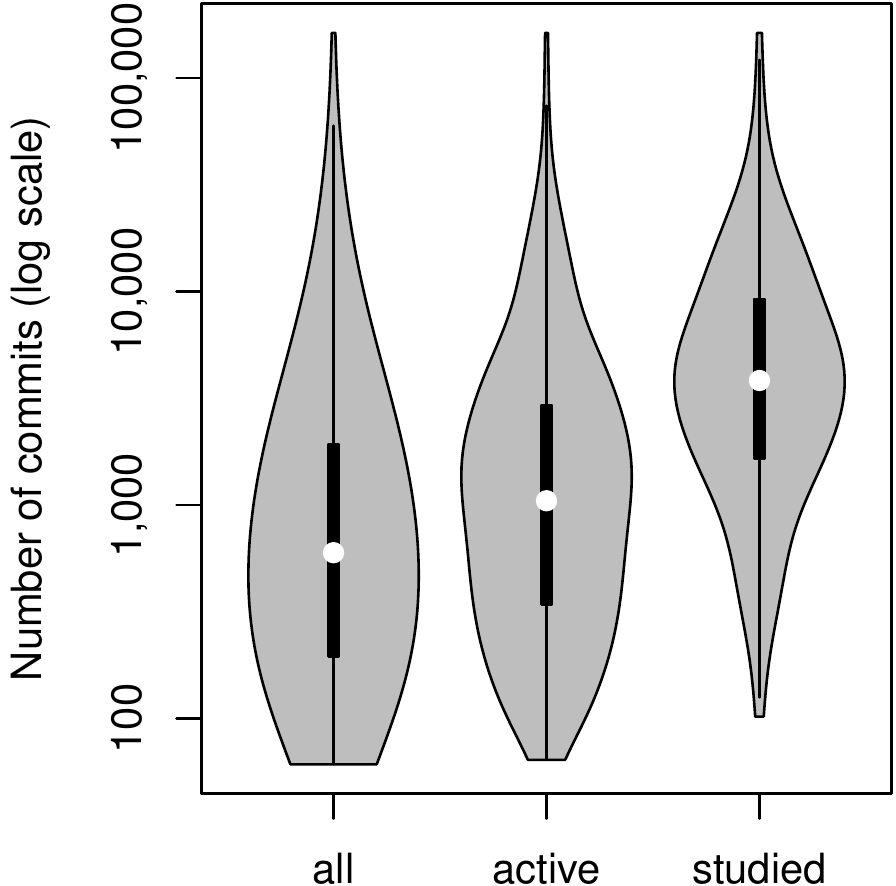}}
\quad
\subfigure[ref2][Source files]{\includegraphics[width=0.232\linewidth]{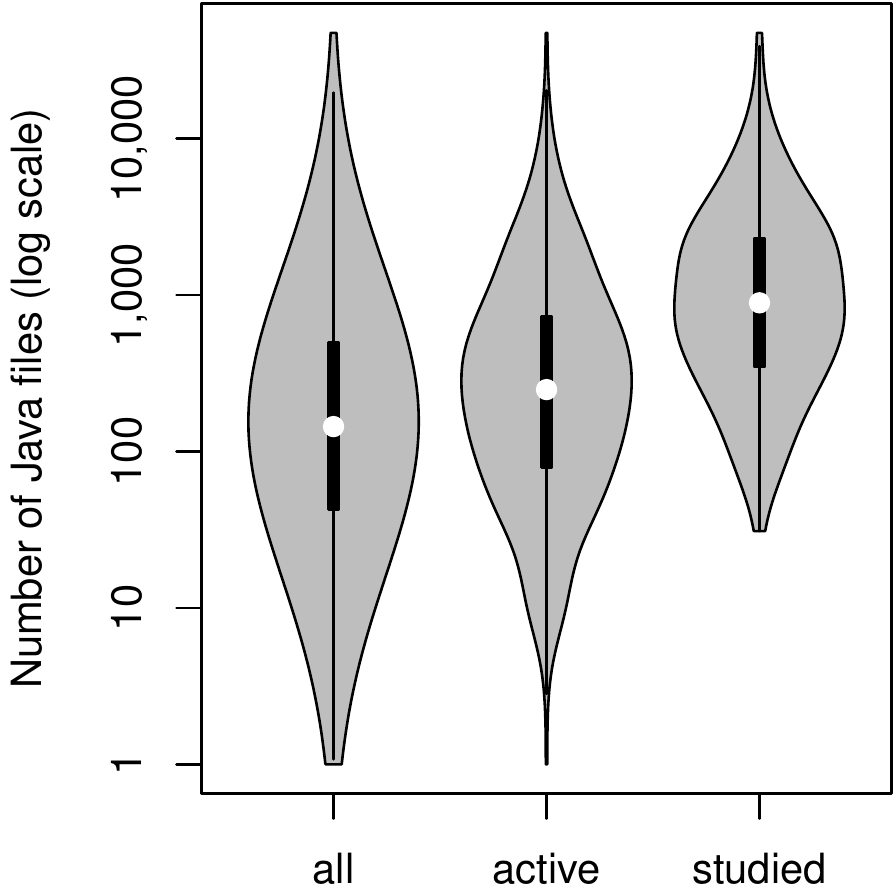}}
\quad
\subfigure[ref2][Contributors]{\includegraphics[width=0.232\linewidth]{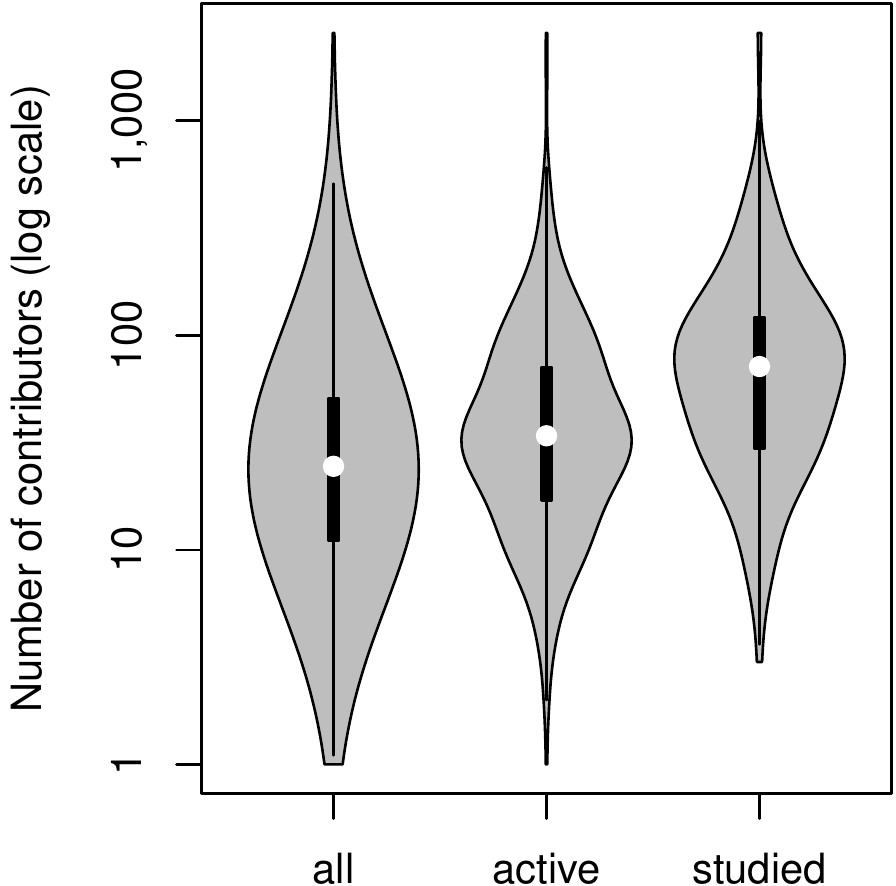}}
\vspace{-5mm}
\caption{Distribution of (a)~age, (b)~commits, (c)~Java source files, and (d)~contributors of repositories}
\vspace{-5mm}
\label{fig:dataset}
\end{figure*}

Figure~\ref{fig:dataset} shows violin plots~\cite{Hintze.Nelson:1998} with the distribution of age (in months), number of commits, size (number of \mcode{*.java} files), and number of contributors of the selected repositories.
We provide plots for all \totalProjects systems (labeled as \textit{all}), for the \activeProjects systems (\activeProjectsPercentage) with at least one commit during the study period (labeled as \textit{active}),
and for the \studiedProjects systems (\studiedProjectsPercentage) effectively analyzed in the study (labeled as \textit{studied}),
which correspond to the repositories with at least one refactoring detected in the commits during the study period (61 days),
along with answers from the developers to our questions about the motivation behind the detected refactorings.
We can observe in Figure~\ref{fig:dataset} that the \textit{active} systems tend to have a higher number of commits, source files, and contributors than the initially selected systems (\textit{all}).
The same holds when comparing the \textit{studied} systems with the \textit{active} systems.
These observations are statistically confirmed by applying the one-tailed variant of the Mann-Whitney \textit{U} test.

\subsection{\toolName Tool}

In the study, we search for refactorings performed in the version history of the selected GitHub repositories by analyzing the differences between the source code of two revisions.
For this purpose, we use a refactoring detection tool proposed in a previous work~\cite{tsantalis_empiricalstudy}.
The tool, named \toolName in this paper, implements a lightweight version of the UMLDiff~\cite{Xing:2005} algorithm for differencing  object-oriented models.
This algorithm is used to infer the set of classes, methods, and fields added, deleted or moved between successive code revisions.
After executing this algorithm, a set of rules is used to identify different types of refactorings.
Unlike other existing refactoring detection tools, such as Ref-Finder~\cite{Kim:2010:RefFinder} and JDevAn~\cite{Xing:2008:JDevAn}, \toolName provides an API and can be used as an external library independently from an IDE,
while Ref-Finder and JDevAn can be executed only within the Eclipse IDE.
The strong dependency of Ref-Finder and JDevAn to the Eclipse IDE prevented us from using these tools in our study,
since as it will be explained in Section~\ref{sec:study_design}, our study required a high degree of automation,
and this could be achieved only by being able to use \toolName programmatically through its API.


In the study, we analyze \refactoringTypes well-known refactoring types detected by \toolName, as listed in the first column of Table~\ref{TabRefactoringActivity}.
The detection of {\sc Rename Class/Method/Field} refactorings is not currently supported by \toolName, because it requires a more advanced source code analysis that examines changes in usage patterns (i.e., changes in class instantiations, method call sites, field accesses, respectively) to verify the consistency of the renaming operation.
Typically, these refactorings are performed to give a more meaningful name to the renamed code element. Previous studies show that they are usually performed automatically, using the refactoring tool support of popular IDEs~\cite{MurphyHill2012, negara2013}.


\subsubsection{\toolName Precision and Recall}
\label{sec:precision_recall}
As we rely on \toolName to find refactorings performed in the version history of software repositories,
it is important to estimate its recall and precision. For this reason, we evaluated \toolName using the dataset reported in a study by Chaparro~et~al.~\cite{Chaparro:2014}.
This dataset includes a list of refactorings performed by two Ph.D. students on two software systems (ArgoUML 
and aTunes) along with the source code before and after the modifications. 
There are 173 refactoring instances in total, from which we selected all 120 instances corresponding to 8 of the refactoring types considered in this study (8 $\times$ 15 instances per type).
The dataset does not contain instances of {\sc Extract Superclass/Interface}, {\sc Move Class}, and {\sc Rename Package} refactorings.
We compared the list of refactorings detected by \toolName with the known refactorings in those systems to 
obtain the results of Table~\ref{TabRefDetectorEval}, which presents the number of true positives (TP), 
the number of false positives (FP), the number of false negatives (FN), the recall and precision for each 
refactoring type.
In total, there are 111 true positives (i.e., existing refactoring instances that were correctly detected) and 9 false negatives (i.e., existing refactoring instances that were not detected), which yield a fairly high recall of 0.93.
Besides, there are 2 false positives (i.e., incorrectly detected refactoring instances), which yield a precision of 0.98.
The lowest observed recall is for {\sc Pull Up Method} (0.80), while the lowest observed precision is for {\sc Extract Method} (0.88).

In conclusion, the accuracy of \toolName is at acceptable levels,
since Ref-Finder (the current state-of-the-art refactoring reconstruction tool)
has an overall precision of 79\% according to the experiments conducted by its own authors~\cite{Prete:2010},
while an independent study by Soares et al.~\cite{Soares:2013} has shown an overall precision of 35\% and an overall recall of 24\% for Ref-Finder.

\begin{table}[htb]
\centering
\renewcommand{\arraystretch}{1.1}
\vspace{-2mm}
\caption{\toolName Recall and Precision}
\label{TabRefDetectorEval}
{\footnotesize
\begin{tabular}{@{}lrrrrr@{}} \toprule
Refactoring & TP & FP & FN & Recall & Prec.\\ \midrule
{\sc Extract Method} & 15 & 2 & 0 & 1.00 & 0.88 \\
{\sc Inline Method}  & 13 & 0 & 2 & 0.87 & 1.00 \\
{\sc Pull Up Attribute}  & 15 & 0 & 0 & 1.00 & 1.00 \\
{\sc Pull Up Method}  & 12 & 0 & 3 & 0.80 & 1.00 \\
{\sc Push Down Attribute}  & 15 & 0 & 0 & 1.00 & 1.00 \\
{\sc Push Down Method}  & 13 & 0 & 2 & 0.87 & 1.00 \\
{\sc Move Attribute}  & 15 & 0 & 0 & 1.00 & 1.00 \\
{\sc Move Method}  & 13 & 0 & 2 & 0.87 & 1.00 \\
\midrule
Total  & 111 & 2 & 9 & 0.93 & 0.98 \\
\bottomrule \end{tabular}
}
\vspace{-5mm}
\end{table}



\subsection{Study Design}
\label{sec:study_design}

During 61 days (between June 8$^{th}$ and August 7$^{th}$ 2015), we monitored all selected repositories to 
detect refactorings.
We built an automated system that periodically fetches commits from each remote repository to a local
copy (using the \texttt{git fetch} operation). Next, the system iterates through each commit and executes \toolName to
find refactorings and store them in a relational database.

As in a previous study~\cite{tsantalis_empiricalstudy}, we compare each examined commit with its parent 
commit in the directed acyclic graph (DAG) that models the commit history in git-based version control repositories.
Furthermore, we exclude merge commits from our analysis to avoid the duplicate report of refactorings.
Suppose that commit $C_M$ merges two branches containing commits $C_A$ and $C_B$, respectively.
Suppose also that a refactoring $\mathit{ref}$ is performed in $C_A$, and therefore detected when we compare 
$C_A$ with its parent commit.
Because the effects of $\mathit{ref}$ are present in the code that resulted from $C_M$, $\mathit{ref}$ 
would be detected again if we compared $C_M$ with $C_B$.
Therefore, we assume that discarding merge commits from our analysis does not lead to any refactoring loss, but rather
avoids duplicate refactoring reports.


On each working day, we retrieved the recent refactorings from the database to perform a manual inspection, using
a web interface we built to aid this task.
In this step, we filter out false positives by analyzing the source code diff of the commit. In this way, 
we avoid asking developers about false refactorings.
Additionally, we also marked commits that already include an explanation for the detected refactoring in
the commit description, to avoid asking an unnecessary question. For instance, in one of the analyzed commits
we found several methods extracted from a method named \texttt{onCreate}, and the commit description was:\margin

\noindent{\em\quotes{Refactored \texttt{AIMSICDDbAdapter::DbHelper\#onCreate} for easier reading}}\margin

Thus, it is clear that the intention of the refactoring was to improve readability by decomposing method 
\texttt{onCreate}. Therefore, it would be unnecessary and inconvenient to ask the developer.

This process was repeated daily, to detect the refactorings as soon as possible after their application in the examined systems.
In this way, we managed to ask the developers shortly after they perform a refactoring, to increase the chances of receiving an accurate response.
We send at most one email to a given developer, i.e., if we detect a refactoring by a developer who has been already contacted before, we do not contact her again, to avoid the perception of our messages as spam email.
The email addresses of the developers were retrieved from the commit metadata.

In each email, we describe the detected refactoring(s) and provide a GitHub URL for the commit where the refactoring(s) is(are) detected.
In the email, we asked two questions:
\begin{enumerate}
\vspace{-1mm}
\item Could you describe why did you perform the listed refactoring(s)? 
\vspace{-2mm}
\item Did you perform the refactoring(s) using the automated refactoring support of your IDE?
\vspace{-1mm}
\end{enumerate}
With the first question, our goal is to reveal the actual motivation behind real refactorings instances.
With the second question, we intend to collect data about the adequacy and usage of refactoring tools, previously investigated in other empirical studies~\cite{MurphyHill2012, negara2013}.
In this way, we can check whether the findings of these studies are reproduced in our study.
We should clarify that by ``automated refactoring'' we refer to user actions that trigger the refactoring engine of an IDE by any means (e.g., through the IDE menus, keyboard shortcuts, or drag-and-drop of source code elements).

During the study period, we sent 465 emails and received 195 responses, achieving a response rate
of 41.9\%. Each response corresponds to a distinct developer and commit.
The achieved response rate is significantly larger than the typical 5\% rate found in questionnaire-based software engineering surveys~\cite{Singer:2008}.
This can be attributed to the \textit{firehouse interview}~\cite{Murphy-Hill:2015} nature of our approach, in which developers provide their feedback
shortly after performing a refactoring and have fresh in their memory the motivation behind it.
Additionally, we included
in our analysis all 27 commits whose description already explained the reasons for the applied refactorings, 
totaling a set of \commitsWithRefactoringAndResponse commits. 
This set of commits covers \studiedProjects different projects and contains \studiedRefactorings refactoring instances in total.

After collecting all responses, we analyzed the answers using thematic analysis~\cite{Cruzes:2011}, a technique for identifying and recording patterns (or \quotes{themes}) within a collection of documents.
Thematic analysis involves the following steps: (1) initial reading of the developer responses, (2) generating initial codes for each response, (3) searching for themes among codes, (4) reviewing the themes to find opportunities for merging, and (5) defining and naming the final themes.
These five steps were performed independently by the first two authors of the paper,  with the support of a simple web interface we built to allow the analysis and tagging of the detected refactorings.
At the time of the study, the first author (Author\#1) had 3 years of research experience on refactoring, while the second author (Author\#2) had over 8 years of research experience on refactoring.

\begin{figure*}[htbp]
	\centering
	\includegraphics[width=0.9\linewidth]{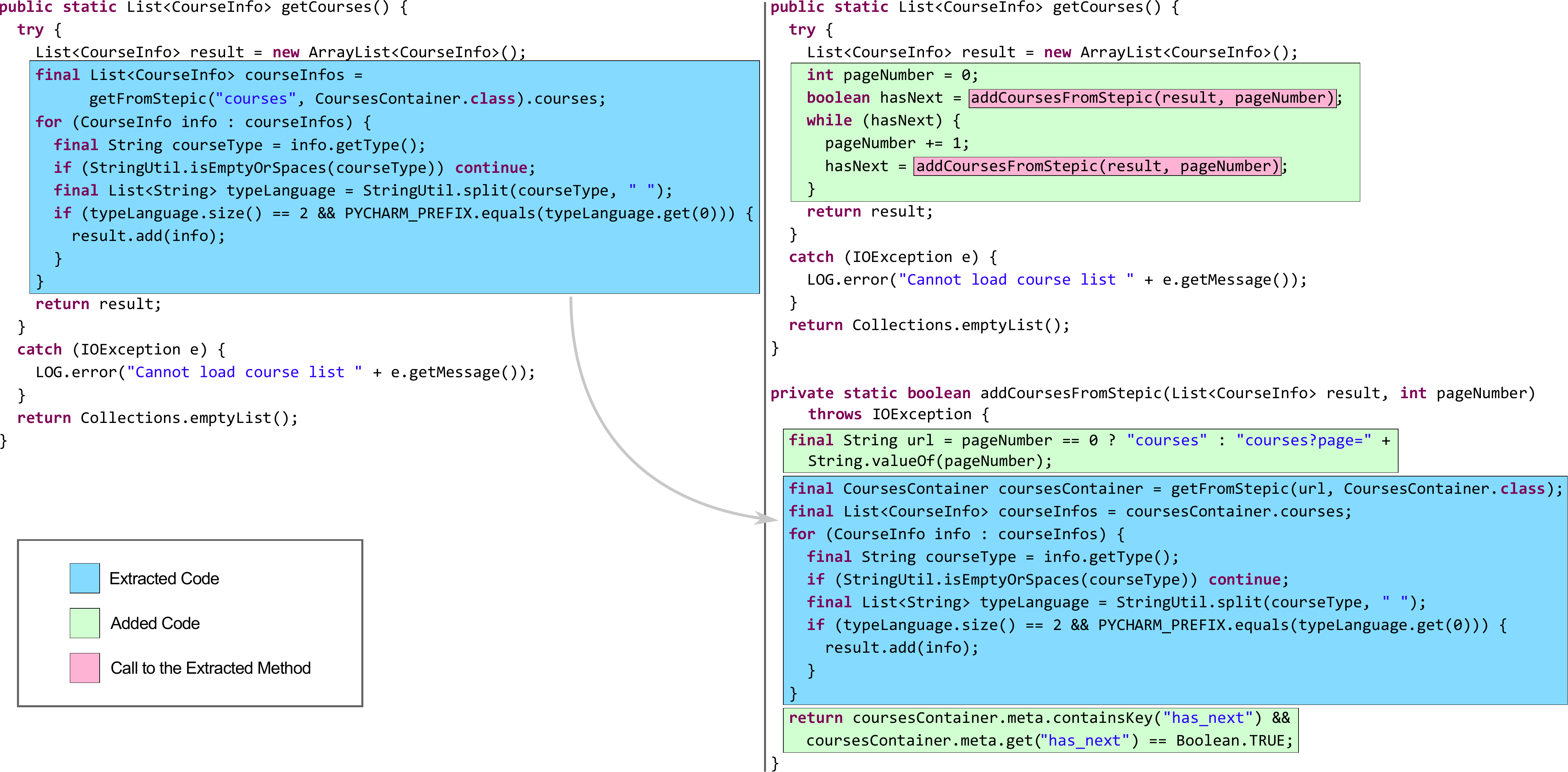}
	\vspace{-4mm}
	\caption{Example of Extract Method refactoring that was required to reach a consensus.}
	\vspace{-4mm}
	\label{fig:consenus_example}
\end{figure*}

After the generation of themes from both authors, a meeting was held to assign the final themes. In 155 cases (58\%), both authors suggested semantically equivalent themes that were rephrased and standardized to compose the final set of themes. The refactorings with divergent themes were then discussed by both authors to reach a consensus. In 94 cases (35\%), one author accepted the theme proposed by the other author. In the remaining 18 cases (7\%), the final theme emerged from the discussion and was different from what both authors previously suggested.
Figure~\ref{fig:consenus_example} shows a case of an {\sc Extract Method} refactoring instance that was required to reach a consensus between the authors. The developer who performed the refactoring explained that the reason for the refactoring was to support a new feature that required pagination, as described in the following comment:\margin

\noindent{\em\quotes{Educational part of PyCharm uses stepic.org courses provid\-er. This server recently decided to use pagination in replies.}}\margin

By inspecting the source code changes, we can see that a part of the original method \texttt{getCourses()} (left-hand side of Figure~\ref{fig:consenus_example})
was extracted into method \texttt{addCoursesFromStepic()} (right-hand side of Figure~\ref{fig:consenus_example}).
After the refactoring, the extracted method is called twice, once before the \texttt{while} loop added in the original method, and once inside the \texttt{while} loop.
For this reason, Author\#1 labeled this case as ``Avoid duplication'', since the extracted method is reused two times after the refactoring.
However, the extracted method contains additional new code to compute properly the URL based on the page number passed as a parameter (first line in the extracted method),
and to return a \texttt{boolean} indicating if there exists a next page (last line in the extracted method).
For this reason, Author\#2 labeled this case as ``Facilitate extension'', since the extracted method also helps to implement the new pagination requirement.
After deliberation, the authors reached a consensus by keeping both theme labels, since the extracted method serves both purposes of reuse and extension.


\subsection{Examined Refactorings}

We monitored \totalProjects Java projects during the study period, and found commits in \activeProjects projects (\activeProjectsPercentage), i.e., 277 projects remained inactive.
We also found \refactoredProjects projects with refactoring activity, as detected by \toolName (including false positives).
In these projects, \detectedRefactorings refactoring instances were detected (in \commitsWithDetectedRefactoring commits), and were manually inspected by the first author of the paper to confirm whether they are indeed true positives.
 
\begin{table}[ht]
\centering
\renewcommand{\arraystretch}{1.1}
\vspace{-5mm}
\caption{Refactoring activity}
\label{TabRefactoringActivity}
{\footnotesize
\begin{tabular}{@{}l@{\hskip 6pt}r@{\hskip 6pt}r@{\hskip 6pt}rr@{\hskip 6pt}r@{}} \toprule
Refactoring & TP & FP & Prec. & Com. & Proj.\\ \midrule
{\sc Extract Method} & 468 & 135 & 0.78 & 312 & 136 \\
{\sc Move Class} & 432 & 512 & 0.46 & 85 & 60 \\
{\sc Move Attribute} & 129 & 44 & 0.75 & 45 & 38 \\
{\sc Rename Package} & 105 & 0 & 1.00 & 25 & 24 \\
{\sc Move Method} & 99 & 48 & 0.67 & 40 & 31 \\
{\sc Inline Method} & 58 & 67 & 0.46 & 44 & 36 \\
{\sc Pull Up Method} & 33 & 1 & 0.97 & 18 & 17 \\
{\sc Pull Up Attribute} & 23 & 1 & 0.96 & 11 & 11 \\
{\sc Extract Superclass} & 22 & 11 & 0.67 & 18 & 16 \\
{\sc Push Down Method} & 16 & 1 & 0.94 & 6 & 6 \\
{\sc Push Down Attribute} & 15 & 1 & 0.94 & 7 & 7 \\
{\sc Extract Interface} & 11 & 8 & 0.58 & 10 & 9 \\
\midrule
Total & 1411 & 830 & 0.63 & 539 & 185 \\
\bottomrule \end{tabular}
}
\vspace{-3mm}
\end{table}

Table~\ref{TabRefactoringActivity} shows the number of true positives (TP), false positives (FP), and precision (Prec.) of \toolName by refactoring type,
as computed after the manual inspection of the detected refactorings.
In general, our tool achieves very high precision for {\sc Rename Package} (100\%), {\sc Pull Up/Push Down Attribute/Method} refactorings (over 94\%),
and relatively high precision for {\sc Extract Method} and {\sc Move Attribute} refactorings (over 75\%),
while the precision for {\sc Move Method} and {\sc Extract Superclass} is 67\%.
However, for some refactorings the precision is closer to 50\%, namely {\sc Extract Interface} (58\%), {\sc Inline Method} (46\%),
and {\sc Move Class} (46\%).
We observed several cases of inner classes falsely detected as moved, because their enclosing class was simply renamed.
By supporting the detection of {\sc Rename Class} refactoring, we could improve {\sc Move Class} precision.
It should be emphasized that we asked the developers only about the true positives detected by \toolName.
In comparison to the results presented in Section~\ref{sec:precision_recall}, the precision is lower,
because the commits analyzed from GitHub projects may include tangled changes, while the commits analyzed in Section~\ref{sec:precision_recall} include only refactoring operations.
Tangled changes make the detection of refactorings more challenging, thus resulting in more false positives.
Finally, Table~\ref{TabRefactoringActivity} shows the number of distinct commits (Com.) and projects (Proj.) 
containing at least one true positive refactoring
(\commitsWithTruePositiveRefactoring out of \commitsWithDetectedRefactoring commits and 185 out of \refactoredProjects projects with detected refactorings contain at least one true positive refactoring).

\section{Why do Developers Refactor?}

\begin{table*}[t]
\centering
\renewcommand{\arraystretch}{1.3}
\caption{{\sc Extract Method} motivations}
\label{TabExtractMethodMotivations}
\footnotesize 
\begin{tabular}{@{}p{0.24\linewidth}p{0.615\linewidth}rl@{}} \toprule
Theme & Description & \multicolumn{2}{c@{}}{Occurrences}\\ \midrule
Extract reusable method & Extract a piece of reusable code from a single place and call the extracted method in multiple places. & 43 & \sbar{43}{43} \\
Introduce alternative method signature & Introduce an alternative signature for an existing method (e.g., with additional or different parameters) and make the original method delegate to the extracted one. & 25 & \sbar{25}{43} \\
Decompose method to improve readability & Extract a piece of code having a distinct functionality into a separate method to make the original method easier to understand. & 21 & \sbar{21}{43} \\
Facilitate extension & Extract a piece of code in a new method to facilitate the implementation of a feature or bug fix, by adding extra code either in the extracted method, or in the original method. & 15 & \sbar{15}{43} \\
Remove duplication & Extract a piece of duplicated code from multiple places, and replace the duplicated code instances with calls to the extracted method. & 14 & \sbar{14}{43} \\
Replace Method preserving backward compatibility & Introduce a new method that replaces an existing one to improve its name or remove unused parameters. The original method is preserved for backward compatibility, it is marked as deprecated, and delegates to the extracted one. & 6 & \sbar{6}{43} \\
Improve testability & Extract a piece of code in a separate method to enable its unit testing in isolation from the rest of the original method. & 6 & \sbar{6}{43} \\
Enable overriding & Extract a piece of code in a separate method to enable subclasses override the extracted behavior with more specialized behavior. & 4 & \sbar{4}{43} \\
Enable recursion & Extract a piece of code to make it a recursive method. & 2 & \sbar{2}{43} \\
Introduce factory method & Extract a constructor call (class instance creation) into a separate method. & 1 & \sbar{1}{43} \\
Introduce async operation & Extract a piece of code in a separate method to make it execute in a thread. & 1 & \sbar{1}{43} \\
\bottomrule \end{tabular}

\vspace{-5mm}
\end{table*}

In this section, we present the results for the first question answered by the developers, regarding the reasons behind 
the application of the refactorings we detected. Based on the results of the thematic analysis process (Section~\ref{sec:study_design}), we compile a catalogue of \totalMotivationThemes distinct motivations.
We dedicate Section~\ref{SecExtractMethodMotivations} to discuss {\sc Extract Method}, which is the most frequently 
occurring refactoring operation in our study, and also the one with the most observed motivations (11). 
Section~\ref{sec:other:motivations} presents the motivations for the remaining refactorings.

\subsection{Motivations for Extract Method}
\label{SecExtractMethodMotivations}
\begin{table*}[!p]
	\centering
	\renewcommand{\arraystretch}{1.3}
	\caption{Motivations for {\sc Move Class, Attribute, Method (MC, MA, MM)}, {\sc Rename Package (RP)} {\sc Inline Method (IM)}, {\sc Extract Superclass, Interface (ES, EI)}, {\sc Pull Up Method, Attribute (PUM, PUA)}, {\sc Push Down Attribute, Method (PDA, PDM)} }
	\label{TabOtherMotivations}
	{\footnotesize
		\begin{tabular}{@{}lp{0.28\linewidth}p{0.52\linewidth}rl@{}} \toprule
Type & Theme & Description & \multicolumn{2}{c@{}}{Occurrences}\\
\midrule
MC & Move class to appropriate container & Move a class to a package that is more functionally or conceptually relevant. & 13 & \sbar{13}{15} \\
MC & Introduce sub-package & Move a group of related classes to a new subpackage. & 7 & \sbar{7}{15} \\
MC & Convert to top-level container & Convert an inner class to a top-level class to broaden its scope. & 4 & \sbar{4}{15} \\
MC & Remove inner classes from deprecated container & Move an inner class out of a class that is marked deprecated or is being removed. & 3 & \sbar{3}{15} \\
MC & Remove from public API & Move a class from a package that contains external API to an internal package, avoiding its unnecessary public exposure. & 2 & \sbar{2}{15} \\
MC & Convert to inner class & Convert a top-level class to an inner class to narrow its scope. & 2 & \sbar{2}{15} \\
MC & Eliminate dependencies & Move a class to another package to eliminate undesired dependencies between modules. & 1 & \sbar{1}{15} \\
MC & Eliminate redundant sub-package & Eliminate a redundant nesting level in the package structure. & 1 & \sbar{1}{15} \\
MC & Backward compatibility & Move a class back to its original package to maintain backward compatibility. & 1 & \sbar{1}{15} \\
\midrule
MA & Move attribute to appropriate class & Move an attribute to a class that is more functionally or conceptually relevant. & 15 & \sbar{15}{15} \\
MA & Remove duplication & Move similar attributes to another class where a single copy of them can be shared, eliminating the duplication. & 4 & \sbar{4}{15} \\
\midrule
RP & Improve package name & Rename a package to better represent its purpose. & 8 & \sbar{8}{15} \\
RP & Enforce naming consistency & Rename a package to conform to project's naming conventions. & 3 & \sbar{3}{15} \\
RP & Move package to appropriate container & Move a package to a parent package that is more functionally or conceptually relevant. & 2 & \sbar{2}{15} \\
\midrule
MM & Move method to appropriate class & Move a method to a class that is more functionally or conceptually relevant. & 8 & \sbar{8}{15} \\
MM & Move method to enable reuse & Move a method to a class that permits its reuse by other classes. & 3 & \sbar{3}{15} \\
MM & Eliminate dependencies & Move a method to eliminate dependencies between classes. & 3 & \sbar{3}{15} \\
MM & Remove duplication & Move similar methods to another class where a single copy of them can be shared, eliminating duplication. & 1 & \sbar{1}{15} \\
MM & Enable overriding & Move a method to permit subclasses to override it. & 1 & \sbar{1}{15} \\
\midrule
IM & Eliminate unnecessary method & Inline and eliminate a method that is unnecessary or has become too trivial after code changes. & 13 & \sbar{13}{15} \\
IM & Caller becomes trivial & Inline and eliminate a method because its caller method has become too trivial after code changes, so that it can absorb the logic of the inlined method without compromising readability. & 2 & \sbar{2}{15} \\
IM & Improve readability & Inline a method because it is easier to understand the code without the method invocation. & 1 & \sbar{1}{15} \\
\midrule
ES & Extract common state/behavior & Introduce a new superclass that contains common state or behavior from its subclasses. & 7 & \sbar{7}{15} \\
ES & Eliminate dependencies & Introduce a new superclass that is decoupled from specific dependencies of a subclass. & 1 & \sbar{1}{15} \\
ES & Decompose class & Extract a superclass from a class that holds many responsibilities. & 1 & \sbar{1}{15} \\
\midrule
PUM & Move up common methods & Move common methods to superclass. & 8 & \sbar{8}{15} \\
\midrule
PUA & Move up common attributes & Move common attributes to superclass. & 7 & \sbar{7}{15} \\
\midrule
EI & Facilitate extension & Introduce an interface to enable different behavior. & 1 & \sbar{1}{15} \\
EI & Enable dependency injection & Introduce an interface to facilitate the use of a dependency injection framework. & 1 & \sbar{1}{15} \\
EI & Eliminate dependencies & Introduce an interface to avoid depending on an existing class/interface. & 1 & \sbar{1}{15} \\
\midrule
PDA & Specialized implementation & Push down an attribute to allow specialization by subclasses. & 2 & \sbar{2}{15} \\
PDA & Eliminate dependencies & Push down attribute to subclass so that the superclass does not depend on a specific type. & 1 & \sbar{1}{15} \\
\midrule
PDM & Specialized implementation & Push down a method to allow specialization by subclasses. & 1 & \sbar{1}{15} \\
\bottomrule \end{tabular}
	}
\vspace{-3mm}
\end{table*}

Table~\ref{TabExtractMethodMotivations} describes 11 motivations for {\sc Extract Method} refactoring and the number of occurrences for each of them. 
The most frequent motivation is to extract a reusable method (43 instances).
In this case, the refactoring is motivated by the immediate reuse of a piece of code
in multiple other places, in addition to the place from which it was originally extracted.
We often observe a concern among developers to reuse code wherever possible, by extracting pieces of reusable code.
This is illustrated by the following comments:\margin

\noindent{\em \quotes{These refactorings were made because of code reusability. I needed to use the same 
code in new method. I always try to reuse code, because when there's a lot of code redundancy it gets
overwhelmingly more complicated to work with the code in future, because when something change in code 
that has it's duplicate somewhere, it usually needs to be changed also there.}}\margin


\noindent{\em \quotes{The reason for me to do the refactoring was: Don't repeat yourself (DRY).}}\margin


The second most frequent motivation is to introduce an alternative method signature for an existing method
 (25 instances), e.g.,~with extra parameters.
To achieve that, the body of the existing method is extracted to a 
new one with an updated signature and additional logic to handle the extended variability.
The original  method is changed to delegate to the new one, passing some default values for the new parameters.
The following comment illustrates this case:\margin

\noindent{\em \quotes{The extracted method \texttt{values(names List<String>, values List<Object>)} could be of 
help for some users using Lists instead of arrays, and because the implementation already transformed the provided 
arrays into Lists internally.}}\margin

Decomposing a method for improving readability (21 instances) is the third most frequent motivation. Typically, 
this corresponds to a {\em Long Method} code smell~\cite{Fowler:1999}, as illustrated in this comment:\margin

\noindent{\em \quotes{The method was so long that it didn't fit onto the screen anymore, so I moved out parts.}}\margin

The next two motivations are to facilitate extension (15 instances) and to remove duplication (14 instances).
In the first case, a method is decomposed to facilitate the implementation of a new feature or the fix of a bug by adding code either in the extracted or in the 
original method, as illustrated in this comment:\margin

\noindent{\em \quotes{I was fixing an exception, in order to do that I had to add the same 
	code to 2 different places. So I extracted initial code, replace 
	duplicate with the extracted method and add the `fix' to the extracted 
	method.}}\margin

In the second case (i.e., remove duplication), a piece of duplicated code is extracted from multiple places into a single method, as illustrated in the following comments:\margin

\noindent{\em \quotes{I refactored shared functionality into a single method.}}\margin

\noindent{\em \quotes{I checked how other test methods create testing User objects and noticed that it takes two lines of code that were repeated all over the test class.
	So I abstracted these two lines of code into a method for better readability and then reused the method in all the places that had the same code.}}\margin

Finally, two other important motivations are to improve testability (6 instances) and to replace a method by preserving backward compatibility (6 instances).
In the first case, the decomposition enables the developer to test parts 
of the code in isolation, as illustrated in this comment:\margin

\noindent{\em \quotes{I wanted to test the part of \texttt{authenticate()} which verifies that a member is element of a set, and that would have been more complex using \texttt{authenticate} directly.}}\margin

In the second case, the goal is to introduce a method having the same functionality with an already existing one, but a different signature (e.g., improved name, or removed unused parameter), and at the same time preserve the public API by making the original method delegate to the new one.
This motivation is best illustrated in the following comment:\margin

\noindent{\em \quotes{I did that refactoring because essentially I wanted to rename the
	functions involved - you'll see the old functions just forward straight to
	the new ones. But I didn't just rename because other code in other projects
	might be referring to the old functions, so they would need to still be
	present (I guess they should have been marked as @deprecated then, but I
	was a bit lazy here).}}

\subsection{Motivations for Other Refactorings}
\label{sec:other:motivations}

Table~\ref{TabOtherMotivations} presents the motivations for the remaining refactorings studied in the paper.
We found nine different motivations for {\sc Move Class}.
The two most frequent motivations are to move a class to a package that is more functionally or conceptually related to 
the purpose of the class (13 instances), and to introduce a sub-package (7 instances). The first one is illustrated by the following comment:\margin


\noindent{\em \quotes{This refactoring was done because common interface for those classes 
lived in \texttt{org.neo4j.kernel.impl.store.\\record}, while most of it's implementors lived in 
\texttt{org.neo4j.\\kernel.impl.store} which did not make sense because all of them are actually records.}}\margin



For {\sc Move Attribute}, the most common motivation is also to move the attribute to an 
appropriate class that is more functionally or conceptually relevant (15 instances), as in the 
example below:\margin

\noindent{\em \quotes{In this case, each of these fields was moved as their relevance changed. As
\texttt{UserService} already handles the login process, it makes sense that changes
to the login process should be encapsulated within \texttt{UserService}.}}\margin

Remove duplication is another motivation for moving an attribute, as illustrated by the following 
comment:\margin

\noindent{\em \quotes{The attributes were duplicated, so I moved them to the proper common place.}}\margin

For {\sc Rename Package}, the most common motivation is to update the name of a package to better represent its purpose (8 instances), as in the example below:\margin

\noindent{\em \quotes{This was a simple package rename. \texttt{test} seems to fit better than \texttt{tests} here as a single test can be executed too.}}\margin

We found three main reasons for a {\sc Move Method} refactoring: move a method to an appropriate 
class (8 instances), move a method to enable reuse (3 instances), and move a method to eliminate 
dependencies (3 instances). 
The most frequent motivation for {\sc Inline Method} is to eliminate 
an unnecessary or trivial method, as illustrated in the comment:\margin

\noindent{\em \quotes{Since the method was a one-liner and was used only in one place, inlining it didn't make 
the code more complex. On the other hand, it allowed to lessen calls to \texttt{getVirtualFile()}.}}\margin

{\sc Extract Superclass} is usually applied to introduce a new class with state or behavior that can be
shared by subclasses (7 instances).
{\sc Pull Up Method/Attribute} is performed to move common code to an existing superclass (8 and 7 instances, 
respectively). {\sc Extract Interface} and {\sc Push Down Attribute/Method} are 
less popular refactorings and thus their motivations have at most two instances.


\section{Refactoring Automation}
\label{SecAutomatedRefactoring}

In this section, we discuss the results drawn from the second question answered by the developers, regarding the 
use (or not) of automatic refactoring tools provided by their IDEs to apply the refactorings we presented. 
First, in Section~\ref{SecRefactoringToolsUnderused}, we present how many of the interviewed developers applied the
refactoring(s) automatically. We also present which refactoring types are more frequently applied with tool support.
In Section~\ref{SecReasonForManualRefactoring}, we discuss some insights drawn from developers' answers that explain
why refactoring is still applied manually in most of the cases.
Last, in Section~\ref{SecIde}, we present additional details regarding which IDE developers most often used
for refactoring.


\subsection{Are refactoring tools underused?}
\label{SecRefactoringToolsUnderused}

Table~\ref{TabManualVsAutomated} shows the results for this question.
95 developers (55\% of valid answers) answered that the refactoring was performed manually without tool support;
66 developers (38\%) answered that the refactoring engine of an IDE was used;
13 developers (7\%) answered that the refactoring was partially automated. 
In summary, refactoring is probably more often applied manually than with refactoring tools.

\begin{table}[htbp]
\centering
\renewcommand{\arraystretch}{1.1}
\vspace{-5mm}
\caption{Manual vs. automated refactoring}
\label{TabManualVsAutomated}
{\footnotesize
\begin{tabular}{@{}p{0.5\linewidth}rl@{}} \toprule
Modification & \multicolumn{2}{c@{}}{Occurrences} \\ \midrule
Manual & 95 & \sbar{95}{95} \\
Automated & 66 & \sbar{66}{95} \\
Not answered & 48 & \sbar{48}{95} \\
Partially automated & 13 & \sbar{13}{95} \\
\bottomrule \end{tabular}
}
\vspace{-3mm}
\end{table}

We also counted the percentage of automated refactorings by refactoring type, as presented in Table~\ref{TabRefactoringTypeAutomated}.
{\sc Rename Package} is the refactoring most often performed with tool support (58\%), followed by {\sc Move Class} (50\%).
Three other refactorings are performed automatically in around a quarter of the cases: {\sc Extract Method} (29\%), {\sc Move Method} (26\%), and {\sc Move Attribute} (24\%).
{\sc Inline Method} follows with 18\% of automatic applications.
Finally, for the remaining refactorings, we do not have a large number of instances to draw safe conclusions (maximum 9 instances),
but there is a consistent trend showing that inheritance-related refactorings are mostly manually applied.

\begin{table}[htbp]
\centering
\renewcommand{\arraystretch}{1.1}
\vspace{-5mm}
\caption{Refactoring automation by type}
\label{TabRefactoringTypeAutomated}
{\footnotesize
\begin{tabular}{@{}lrlrl@{}} \toprule
Refactoring Type & \multicolumn{2}{c}{Occurrences} & \multicolumn{2}{c@{}}{Automated \%} \\ \midrule
{\sc Extract Method} & 118 & \sbar{118}{118} & 29\% & \sbar{34}{118} \\
{\sc Move Class} & 36 & \sbar{36}{118} & 50\% & \sbar{18}{36} \\
{\sc Move Attribute} & 21 & \sbar{21}{118} & 24\% & \sbar{5}{21} \\
{\sc Move Method} & 19 & \sbar{19}{118} & 26\% & \sbar{5}{19} \\
{\sc Inline Method} & 17 & \sbar{17}{118} & 18\% & \sbar{3}{17} \\
{\sc Rename Package} & 12 & \sbar{12}{118} & 58\% & \sbar{7}{12} \\
{\sc Extract Superclass} & 9 & \sbar{9}{118} & 11\% & \sbar{1}{9} \\
{\sc Pull Up Method} & 9 & \sbar{9}{118} & 11\% & \sbar{1}{9} \\
{\sc Pull Up Attribute} & 7 & \sbar{7}{118} & 14\% & \sbar{1}{7} \\
{\sc Extract Interface} & 3 & \sbar{3}{118} & 0\% & \sbar{0}{3} \\
{\sc Push Down Attribute} & 3 & \sbar{3}{118} & 33\% & \sbar{1}{3} \\
{\sc Push Down Method} & 2 & \sbar{2}{118} & 0\% & \sbar{0}{2} \\
\bottomrule \end{tabular}
}
\vspace{-3mm}
\end{table}

\subsection{Why do developers refactor manually?}
\label{SecReasonForManualRefactoring}

29 developers explained in their answers why they did not use a refactoring tool. 
Table~\ref{TabManualRefactoringReasons} shows five distinct themes we identified in these answers.

\begin{table}[htbp]
\centering
\renewcommand{\arraystretch}{1.3}
\vspace{-4mm}
\caption{Reasons for not using refactoring tools}
\label{TabManualRefactoringReasons}
{\footnotesize
\begin{tabular}{@{}p{0.74\linewidth}rl@{}} \toprule
Description & \multicolumn{2}{c@{}}{Occurrences}\\ \midrule
The developer does not trust automated support for complex refactorings. & 10 & \sbar{10}{10} \\
Automated refactoring is unnecessary, because the refactoring is trivial and can be manually applied. & 8 & \sbar{8}{10} \\
The required modification is not supported by the IDE. & 6 & \sbar{6}{10} \\
The developer is not familiar with the refactoring capabilities of his/her IDE. & 3 & \sbar{3}{10} \\
The developer did not realize at the moment of the refactoring that he/she could have used refactoring tools.
& 2 & \sbar{2}{10} \\
\bottomrule \end{tabular}
}
\end{table}

Lack of trust (10 instances) was the most frequent reason. Some developers do not trust
refactoring tools for complex operations that involve code manipulation and only use them for renaming
or moving:\margin

\noindent{\em \quotes{I don't trust the IDE for things like this, and usually lose other comments, notation, 
spacing from adjacent areas.}}\margin

\noindent{\em \quotes{I'd say developers are reluctant to let a tool perform anything but trivial refactorings, 
such as the ones you picked up on my commit.}}\margin

On the other hand, some developers also think that tool support is unnecessary in simple cases (8 instances).
Sometimes the operation may involve only local changes and is trivial to do by hand. Thus, calling a 
special operation to do it is considered unnecessary, as illustrated by this comment:\margin

\noindent{\em \quotes{Automated refactoring is overkill for moving some private fields.}}\margin

Additionally, developers also mentioned: lack of tool support for the specific refactoring they were 
doing (6 instances), not being familiar with refactoring features of the IDE (3~instances), and not
realizing they could use refactoring tools at the moment of the refactoring (2~instances).

\subsection{What IDEs developers use for refactoring?}
\label{SecIde}

When answering to our emails, 83 developers spon\-ta\-ne\-ously mentioned which IDE they use. 
Therefore, we decided to investigate these answers, specially because our study is not dependent
on any IDE, and thus differs from previous studies which are usually based only on 
Eclipse data~\cite{MurphyHill2012,negara2013}.
Table~\ref{TabIdePopularity} shows the most common IDEs mentioned in these answers and the percentage 
of refactorings performed automatically in these cases.
139 developers (63\%) did not explicitly mention an IDE when answering this question. 
Considering the answers citing an IDE, IntelliJ IDEA is the most popular one. 
It also has the highest ratio of refactorings performed automatically (71\%). 
Since 11 {\sc JetBrains/\-intellij-community} (and related plug-ins) developers answered to our questions, 
we also investigated the answers separately in two groups, namely 
answers from IntelliJ IDEA developers and from IntelliJ IDEA users.
We observed that the ratio of automated refactorings in both groups is very similar (73\% vs. 70\%).
Therefore, the responses from these 11 IntelliJ IDEA developers do not bias the percentage of automated refactoring reported for 
IntelliJ IDEA. 



\begin{table}[htbp]
\centering
\renewcommand{\arraystretch}{1.1}
\vspace{-5mm}
\caption{IDE popularity}
\label{TabIdePopularity}
{\footnotesize
\begin{tabular}{@{}lrlrl@{}} \toprule
IDE & \multicolumn{2}{c}{Occurrences} & \multicolumn{2}{c@{}}{Automated \%} \\ \midrule
Editor not mentioned & 139 & \sbar{139}{139} & 12\% & \sbar{17}{139} \\
IntelliJ IDEA & 51 & \sbar{51}{139} & 71\% & \sbar{36}{51} \\
Eclipse & 18 & \sbar{18}{139} & 44\% & \sbar{8}{18} \\
NetBeans & 8 & \sbar{8}{139} & 50\% & \sbar{4}{8} \\
Android Studio & 4 & \sbar{4}{139} & 25\% & \sbar{1}{4} \\
Text Editor & 2 & \sbar{2}{139} & 0\% & \sbar{0}{2} \\
\bottomrule \end{tabular}
}
\vspace{-3mm}
\end{table}


\section{Discussion}

In this section, we discuss the main findings of our study.\margin


\noindent{\bf Refactoring Motivations:} Our study confirms that {\sc Extract Method} is  the 
``Swiss army knife of refactorings''~\cite{tsantalis_empiricalstudy}. It is the refactoring with the most motivations (11 in total).
In comparison to~\cite{tsantalis_empiricalstudy}, there is an overlap in the reported motivation themes for {\sc Extract Method}.
We found some new themes, such as \emph{improve testability} and \emph{enable recursion}, but we did not find any instances of
the themes \emph{encapsulate field} and \emph{hide message chain}, reported in~\cite{tsantalis_empiricalstudy}, which are related to code smell resolution.
We assume these different themes are due to the nature of the examined projects, since~\cite{tsantalis_empiricalstudy} examined only
three libraries and frameworks, while in this study we examined \studiedProjects projects from various domains including standalone applications.
By comparing to the code symptoms that initiate refactoring reported in the study by Kim et al.~\cite{kim-tse-2014}, we found the \emph{readability},
\emph{reuse}, \emph{testability}, \emph{duplication}, and \emph{dependency} motivation themes in common.

Most of the refactoring motivations we found have the intention to facilitate or even enable the completion of the maintenance task that the developer is working on.
For instance, \emph{extract reusable method}, \emph{introduce alternative method signature}, and \emph{facilitate extension}
are among the most frequent motivations, and all of them involve enhancing the functionality of the system.
Therefore, {\sc Extract Method} is a key operation to complete other
maintenance tasks, such as adding a feature or fixing a bug.
In contrast, only two out of the 11 motivations we found (\emph{decompose method to improve readability} and \emph{remove duplication}) are targeting code smells. 
This finding could motivate researchers and tool builders to design
refactoring recommendation systems~\cite{Tsantalis:2011, Silva:2014, Tairas:2012, Hotta:2012, Meng:2015, Tsantalis:2015}
that do not focus solely on detecting refactoring opportunities for the sake of code smell resolution, but can support other refactoring motivations as well.

We also observe that developers are seriously concerned about avoiding code duplication, when working on a given maintenance task. They often use refactorings---especially {\sc Extract Method}---to
achieve this goal, as illustrated by the following comments:\margin

\noindent{\em \quotes{I need to add a check to both the then- and the else-part of an if-statement. 
This resulted in more duplicated code than I was comfortable with.}}\margin

\noindent{\em \quotes{There was already code duplication, but the bug fix required another cut-and-paste, 
which made it code triplication. That was above my pain level so I decided to group the replicated code 
out into \texttt{bail()}.}}\margin

The other refactorings we analyzed are typically performed to improve the system design. For example,
the most common motivation for {\sc Move Class}, {\sc Move Attribute}, and {\sc Move Method} is to reorganize
code elements, so that they have a stronger functional or conceptual relevance.\margin

\noindent{\bf Automated vs. Manual Refactoring:} In a field study with Eclipse users, 
Negara et al.~\cite{negara2013} report that most refactorings (52\%) are manually performed.
In our study, involving developers using a wider variety of IDEs, we found that 55\% of refactorings are manually performed. 
However, we also found that IntelliJ IDEA users tend to use more the refactoring tool support than other IDE users.
Moreover, the results for automated {\sc Extract Method} refactorings are very similar in both studies: 28\% in our study, 
against 30\% in their study.
While the total percentages of manually performed refactorings are very similar, we should
keep in mind that Negara et al.~counted simple refactorings, like renamings, which are more often applied with tool support.
Compared to the study by Murphy-Hill et al.~\cite{MurphyHill2012},
where they report that 89\% of refactorings are performed manually (considering also renamings),
we detected significantly more automated refactorings.
We suspect this difference may be due to two reasons.
First, automated refactoring tools may have become more popular and reliable
over the last years. Second, our study involves developers using a broader range of IDEs, which may also 
influence how developers use refactoring tool support.

Regarding the reasons for not using automated refactoring, our results are in line with the three main factors found in the study by 
Murphy-Hill et al.~\cite{MurphyHill2012}: \emph{awareness}, \emph{opportunity}, and \emph{trust}. 
The exception is the argument that tool support is unnecessary in simple cases, which is not closely 
related to any of the three aforementioned factors.
However, the same argument can be observed in the study by Kim~et~al.~\cite{kim-tse-2014}, in which some 
developers mention that they do not feel a great need for automated refactoring tools.\margin

\noindent{\bf Refactoring Popularity:}
In this study we detected refactorings in \refactoredProjects 
of the monitored repositories in a time window of 61 days. Given that only \activeProjects out of the \totalProjects monitored 
repositories were active during that period, we found refactoring activity in 60.5\% of the repositories with at least
one commit. This shows that refactoring is a common practice, especially considering that 
frequent refactorings such as {\sc Rename Class/Method/Field} were not considered.
\begin{comment}
It is also worth noting that only in 22\% of the commits containing refactoring the
developer mentioned it in the commit description. Thus, relying in such criterion to find refactorings 
would have been a serious limitation, as also suggested by the study of Murphy-Hill et al.~\cite{MurphyHill2012}.
\end{comment}

The top-5 most popular refactorings detected in our study are 
{\sc Extract Method}, {\sc Move Class}, {\sc Move Attribute}, {\sc Rename Package}, and {\sc Move Method}.
{\sc Move Method} is the third most popular refactoring in the study by Negara et al.~\cite{negara2013}.
The top-2 refactorings in this study ({\sc Rename Local Variable} and {\sc Extract Local Variable}) are low-level refactorings, which have not been considered in our study.
We focused on high-level refactorings, because they can be motivated by multiple factors.

Using a sample of 40 commits with manual and automated refactorings, Murphy-Hill 
et al.~\cite{MurphyHill2012} report that the two most popular refactorings are {\sc Rename Constant} and {\sc Push Down}. 
However, {\sc Push Down} refactorings are among the least popular ones in our study. This difference 
may be related to the number of commits analyzed in the studies (40 vs. \commitsWithTruePositiveRefactoring commits in our study),
and the specialized nature of the software (i.e., the Eclipse IDE) examined in \cite{MurphyHill2012}.

\section{Threats to Validity}

\noindent \textbf{External Validity}: This study is restricted to open source, Java-based, GitHub-hosted projects. Thus, 
we cannot claim that our findings apply to industrial systems, or to systems implemented in other programming languages.
However, we collected responses from \studiedDevelopers developers contributing in \studiedProjects different projects, which is one of the largest samples
of systems used in refactoring studies.\margin

\noindent \textbf{Internal Validity}: First, we use in the study a tool that detects refactorings by comparing 
two revisions of the code.
We evaluated the recall of this tool using a sample of 120 documented refactoring operations. We achieved a recall of 0.93.
However, we cannot guarantee a similar recall in the studied GitHub projects, because some commits might contain tangled changes making more difficult to isolate (or untangle \cite{Dias:2015}) the changes related to refactorings.
In addition, it is known that this kind of detection approach may miss refactorings that do not reach the 
version control system (e.g., sequences of overlapping refactorings applied to the same piece of code).
We claim this threat should be tolerated in large scale studies, where we cannot assume that the developers would be willing to install an external monitoring tool in their IDEs~\cite{negara2013}.
Furthermore, as we showed in this study, developers nowadays use IDEs from multiple vendors.
In order to cover as many IDEs as possible and strengthen the external validity,
a study based on monitoring would require to develop a separate version of this tool for each IDE.
Second, we cannot claim that the catalogue of motivations we propose is exhaustive.
Notably, we have a limited number of motivation themes 
for less frequent refactoring types, such as {\sc Push Down Method/Attribute} and {\sc Extract 
Interface}.
Third, to mitigate inconsistencies in the proposed themes, we rely on an initial classification
performed independently by two authors of the paper, followed by a consensus building process.
We also make publicly available the responses collected from the developers and the proposed
refactoring motivation themes to provide a means for replication and verification.

%


\section{Conclusions}
In summary, the main conclusions and lessons learned are:

\begin{enumerate}[leftmargin=*]
	\setlength\itemsep{0mm}
	\vspace{-2.5mm}
	\item Refactoring activity is mainly driven by changes in the requirements (i.e., new feature and bug fix requests)
	and much less by code smell resolution.
	Only 2 out of the 11 motivations for {\sc Extract Method} were related to code smell resolution (\emph{remove duplication}, \emph{decompose method}) covering only 25\% (35/138) of the motivation instances.
	\vspace{-1.3mm}
	\item {\sc Extract Method} is a key operation that serves multiple purposes, specially those related to code
	reuse and functionality extension. It is also used to improve the testability of code, and deprecate public API methods.
	\vspace{-1.3mm}
	\item The elimination of dependencies is the most common motivation among the \emph{move}/\emph{abstract} related refactorings.
	\vspace{-1.3mm}
	\item Manual refactoring is still prevalent (55\% of the developers refactored manually the code).
	In particular, inheritance related refactoring tool support seems to be the most under-used (only 10\% done automatically),
	while {\sc Move Class} and {\sc Rename Package} are the most trusted refactorings (over 50\% done automatically).
	\vspace{-1.3mm}
	\item The IDE plays an important role in the adoption of refactoring tool support. IntelliJ IDEA users perform more automated refactorings (71\% done automatically) than Eclipse users (44\%) and Netbeans users (50\%).
	\vspace{-1.3mm}
	\item Compared to the study by Murphy-Hill et al.~\cite{MurphyHill2012}, it seems that developers apply more automated refactorings nowadays.
	Our findings confirm Negara et al.~\cite{negara2013} who collected data only from Eclipse IDE users, but our study covers developers using more IDEs.
	\vspace{-2mm}
\end{enumerate}

Based on our findings, we propose that future research on refactoring recommendation systems should refocus from 
code-smell-oriented to maintenance-task-oriented solutions.
This could be achieved by leveraging the recent advancements in feature location~\cite{Dit:2013} and requirements tracing to automatically locate the code
associated with a feature or bug fix request, or a requirement change, and recommend suitable refactorings that will make easier the completion
of the maintenance task.
We strongly believe this will boost the adoption of recommendation systems by the developers.

\section*{Acknowledgments}
Danilo Silva and Marco Tulio Valente are partially funded by FAPEMIG and CNPq.
Nikolaos Tsantalis is partially funded by NSERC and FRQNT.
We gratefully thank the \studiedDevelopers developers who participated in our study.

\bibliographystyle{abbrv}
\bibliography{refactoring-study}  

\end{document}